# Multiple front propagation into unstable states

R. Montagne, A. Amengual, and E. Hernández-García

*Departament de Física*
*Universitat de les Illes Balears*
*E-07071 Palma de Mallorca (Spain)*

M. San Miguel*

*Department of Mathematics*
*University of Arizona*
*Tucson, Arizona 85721 (USA)*
(December 10, 1993)

The dynamics of transient patterns formed by front propagation in extended nonequilibrium systems is considered. Under certain circumstances, the state left behind a front propagating into an unstable homogeneous state can be an unstable periodic pattern. It is found by a numerical solution of a model of the Fréedericksz transition in nematic liquid crystals that the mechanism of decay of such periodic unstable states is the propagation of a second front which replaces the unstable pattern by a another unstable periodic state with larger wavelength. The speed of this second front and the periodicity of the new state are analytically calculated with a generalization of the marginal stability formalism suited to the study of front propagation into periodic unstable states.

## I. INTRODUCTION

Front propagation is a common phenomenon in systems driven out from equilibrium [1]. It can occur when different regions of an extended system are prepared into unequivalent states. The interface between both regions is often sharp and moves at a characteristic velocity. In potential systems, for which the governing equations can be derived from a Lyapunov functional, the relative stability of the different states determines the direction in which the front moves. In nonpotential systems there are different stability criteria from which different kinds of fronts and front bifurcations can be obtained [2–4]. Interesting cases of patterns formed by chemical front propagation have been recently reported [5].

Early rigorous work in front propagation addressed the advance of a uniform stable state into a uniform unstable one [6,7]. An important step forward was the recognition [8,9] that the rigorous results for the velocity of the front could be recovered from a simple marginal stability criterion [10–12]. It basically consists in studying the stability of the moving front neasecurityr its leading edge and stating that the selected velocity is the smallest among the ones for which the front is linearly stable. The physics justifying such prescription is that in the usual situation the state of the system accelerates along the manifold of unstable fronts until the first stable one is reached [13]. The marginal stability criterion cannot be applied when a nonlinear front dominates over the linear one [12,13], but it gives correct answers in many circumstances including nonpotential situations. The relation of the criterion with a structural principle has been recently addressed [14].

To apply the marginal stability prescription for front propagation into an homogeneous state, one writes the solution $\psi(x,t)$ of the relevant equations as $\psi(x,t) \sim \exp(iqx + \omega(q)t)$, linearize in $\psi$, and find the dispersion relation $\omega(q)$. The marginally stable front velocity $v^*$ is obtained from the simultaneous solution $(q^*, v^*)$ of

$$v^* = \frac{\text{Re}[\omega(q^*)]}{\text{Im}[q^*]} = -\text{Im}\left[\frac{d\omega(q)}{dq}\bigg|_{q^*}\right], \quad \text{Re}\left[\frac{d\omega}{dq}\bigg|_{q^*}\right] = 0, \tag{1.1}$$

where $q$ and $q^*$ should be considered as complex. This criterion has been successfully applied to situations where no exact results were available, such as vortex-front propagation in rotating Couette-Taylor flow [15,16] and Rayleigh-Bénard convection [17]. Note that in these cases $\text{Re}[q^*] \neq 0$ and the stable state that advances into the unstable one is not homogeneous, but periodic. In this situation, front propagation has additional importance since it provides a natural mechanism for pattern selection. The marginal stability hypothesis, with some additional assumptions [8], gives the wavelength $\lambda$ of the pattern left behind the front:

$$\lambda^{-1} = \frac{1}{2\pi} \left( \frac{\text{Im}\left[\omega(q^*)\right]}{v^*} + \text{Re}\left[q^*\right] \right) \ . \tag{1.2}$$

Front propagation can also produce unstable periodic patterns. This was explicitly demonstrated in [18] for the extended Fisher-Kolmogorov model, an equation for which the only stable states are homogeneous. When this system is prepared in an initial condition with a sufficiently sharp interface separating the stable homogeneous state from an unstable homogeneous region, the advance of the front leads to the formation of a periodic pattern. This phenomenon could be considered as a generalization of the production of metastable phases by front propagation [19]. The situation is reminiscent of others in which a periodic pattern that is not stable appears after a linear instability [20]. At late times, the transient pattern evolves and decays.

In the context of a model for the dynamics of nematic liquid crystals (which includes as a particular case the Cahn-Hilliard equation of spinodal decomposition) we find here a new process by which the unstable periodic pattern created by the propagating front approaches the homogeneous stable state: a second front advances through the periodic unstable pattern leaving behind a new periodic state with a larger wavelength. Under certain circumstances, the phenomenon can repeat itself until the homogenous state is attained. Our main goal in this paper is to study the characteristics of this second front and of the pattern created. We calculate the velocity of the second front and the periodicity of the pattern behind it. This is done by adapting the marginal stability theory to this new situation. The resulting equations are approximately solved and compared with numerical simulations.

Front propagation into periodic unstable states has been discussed previously in [21] and [22]. In [21] the propagating front solutions of the 'amplitude equation' are considered with periodic unstable solutions as initial conditions. The difference with our situation is that in an amplitude equation description such initial condition can not be created by a primary front, since the front leaves a stable periodic pattern. In addition the decay of an initial unstable pattern is to a final stable pattern. More related to our case is the situation of [22], where complicated sequences of fronts and front splitting were found. A major technical difference between our situation and the ones discussed [21,22] is that the linear analysis around their stationary solutions lead to equations with constant coefficients, so that the usual formalism of marginal stability theory can be applied directly. We will deal here with linear equations with periodic coefficients, which is the generic case in the stability analysis around periodic solutions. Thus we need some adaptation of the marginal stability analysis to our case.

The paper is organized as follows: In section II some general aspects of models leading to unstable patterns are reviewed. In section III, a marginal stability theory to characterize propagating fronts into an unstable periodic state is set-up. This theory is applied in section IV to a particular model realization for the Fréedericksz transition in nematics. In section V, the velocity of the second front obtained from such theory and the wavelength of the pattern left behind are compared with values from the numerical solution of the model system, and the subsequent evolution of the pattern is discussed qualitatively from the observed simulations. Finally, section VI is devoted to conclusions.

## II. MODELS LEADING TO TRANSIENT PATTERNS

In the general context of front propagation in potential systems, several cases can occur. The front separates states of different relative stability. The 'less stable' state can be either unstable or metastable and the 'more stable' can be again unstable or metastable or, obviously, stable. More important from the point of view of this paper, each state can be homogeneous or periodic. It is possible to find in the literature several examples of some of these cases. In the work of [18,11], unstable periodic patterns were shown to be formed behind a propagating front in the extended Fisher-Kolmogorov model (EFK). The EFK model is defined by the following equation for a scalar variable $\psi(x,t)$:

$$\dot{\psi}(x,t) = \partial_x{}^2 \psi - \gamma \partial_x{}^4 \psi + c\psi - b\psi^3 \ . \tag{2.1}$$

This is a generalization of the original Fisher-Kolmogorov (FK) equation (the model A of critical dynamics when a noise term is introduced) which is Eq.(2.1) with $\gamma = 0$.

In the context of the nematodynamic equations for liquid crystals, a model for the analysis of some general aspects of transient pattern evolution was analyzed in a recent paper [23]:

$$\dot{\psi}(x,t) = (a - \partial_x^2)(\partial_x^2 \psi + c\psi - b\psi^3) \tag{2.2}$$

Eq. (2.2) defines a potential model which will be called Modified Cahn-Hilliard model (MCH). The original Cahn-Hilliard equation (CH) is the particular case of (2.2) with $a = 0$. The MCH model has also been considered by Salje [24] in a different context, for a partially conserved dynamics. It has also been shown to form unstable patterns by front propagation [25]. In the rest of the paper, we will assume $a, \gamma, c, b > 0$. When numerical values are needed, we take $c = 1$ without loss of generality, and $b = 3$, as appropriate for modeling the Fréedericksz transition in nematics with model MCH [23]. The free parameters of our study will be $\gamma$, and specially $a$.

Let us summarize some of the properties of the four models defined so far (FK, EFK, MCH, and CH): All of them have in common the existence of at least three kinds of bounded stationary solutions: (a) $\psi(x) = \pm\sqrt{c/b} \equiv \psi_{\pm}$, (b) $\psi(x) = 0$, and (c) a family of periodic solutions $\psi_q(x)$ of fundamental wavenumber $q$ in the range $(0, q_{max})$ [23]. For the EFK model, $q_{max} = ((\sqrt{1+4\gamma c} - 1)/2\gamma)^{\frac{1}{2}}$, and $q_{max} = \sqrt{c}$ in the other models. For periodic boundary conditions this completes the list of stationary solutions for the MCH and FK models. There are additional solutions of the CH equation and EFK model not considered here.

Regarding stability with respect to small perturbations, both solutions included in type (a) are linearly stable, and solution (b) is linearly unstable. All the periodic stationary solutions (c) are also unstable [23].

When $a > c$, there is a change of variables transforming the linear part of the MCH model into the linear part of the EFK. Since marginal stability theory makes use only of the linear portion of the evolution equations, the results obtained in [18] for the fronts in the EFK model can be immediately translated to the MCH model for $a > c$. The mapping is obtained by rescaling $x$ and $t$ in the MCH model into $x' = x\sqrt{a/(a-c)}$ and $t' = t/a$. Then the EFK model is obtained in the new variables with

$$\gamma = \frac{a}{(a-c)^2} \qquad (2.3)$$

Fronts advancing into the unstable uniform state in the EFK model leave behind one of the equilibrium homogeneous states $\psi_{\pm}$ if $\gamma > 1/12$, and a periodic pattern if $\gamma < 1/12$ [18]. In the MCH model this last region is mapped into $a < c(7 + 4\sqrt{3}) \approx 13.93c$.

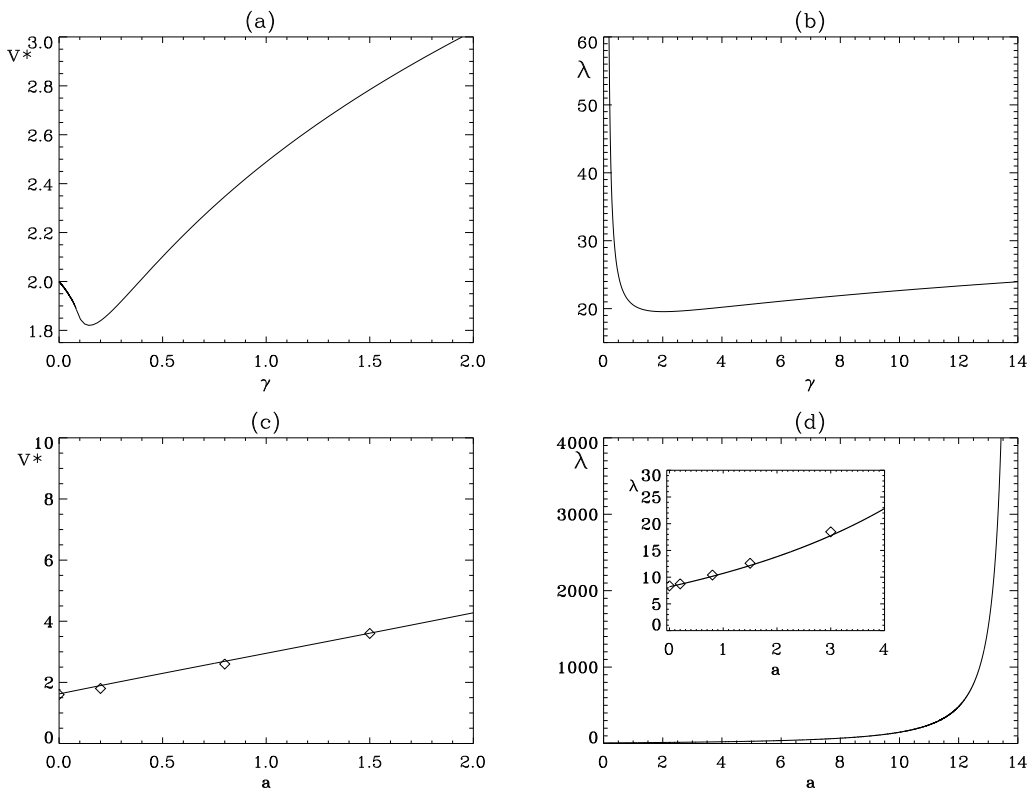

FIG. 1. Velocity of the first front calculated from (1.1) and wavenumber of the pattern left in the EFK model (a and b) and in the MCH model (d and e). Diamonds are simulation results for MCH model.

For $a < c$, the EFK and MCH can not be mapped into each other and the marginal stability theory should be applied directly using the linear dispersion relation for the MCH model, $\omega(q) = (a + q^2)(c - q^2)$, in Eq. (1.1). The procedure does not present any new difficulty. In Fig. 1 we show the velocity of the front moving into the homogeneous state and the wavenumber of the pattern left behind in the EFK model (as a function of $\gamma$) and in the MCH model (as a function of $a$ with $c = 1$). For $a < c$, both the wavelength of the pattern and the velocity of the front in the MCH model are small, hence a large amount of periods can be obtained with a small front velocity. This circumstance never happens in the EFK model. We also remark that it is precisely $a < c$ the situation for which the MCH model gives results qualitatively different from the FK model [23] and appropriate to describe the Fréedericksz transition in nematic liquid crystals.

In Fig. 2 we show front propagation from a numerical solution of the MCH model for $a = 0.2$. Some algorithmic details follow: The integration of Eq.(2.2) has been performed by using a centered finite-difference scheme up to order $(dx)^4$ to approximate the spatial derivatives. To determine $\psi$ at $t + dt$ a single step predictor-corrector method is used with $dx = 0.25$ and $dt = 10^{-4}$ [23]. The initial condition was $\psi(x, t = 0) = 1/\sqrt{b}$ for $x < l$ and $\psi(x, t = 0) = 0$ for $x > l$, $l$ being a suitable distance (for instance, $l = 20$). The boundary condition on the left was $\psi(x < 0, t) = 1/\sqrt{b}$. The size $L$ of the system is initially large enough ($L = 500$) so that when the front arrives to the end, there is a well developed periodic structure on the left. Once the front arrives to the right end, the size of the system is slightly reduced by taking only the left region, and the remaining right part is taken as a new boundary condition, forced to be stationary from now on. By using larger systems, we have checked that the evolution at some distance from the right boundary is independent of this particular prescription for the boundary condition, as we expected from the local nature of the evolution discussed in [23].

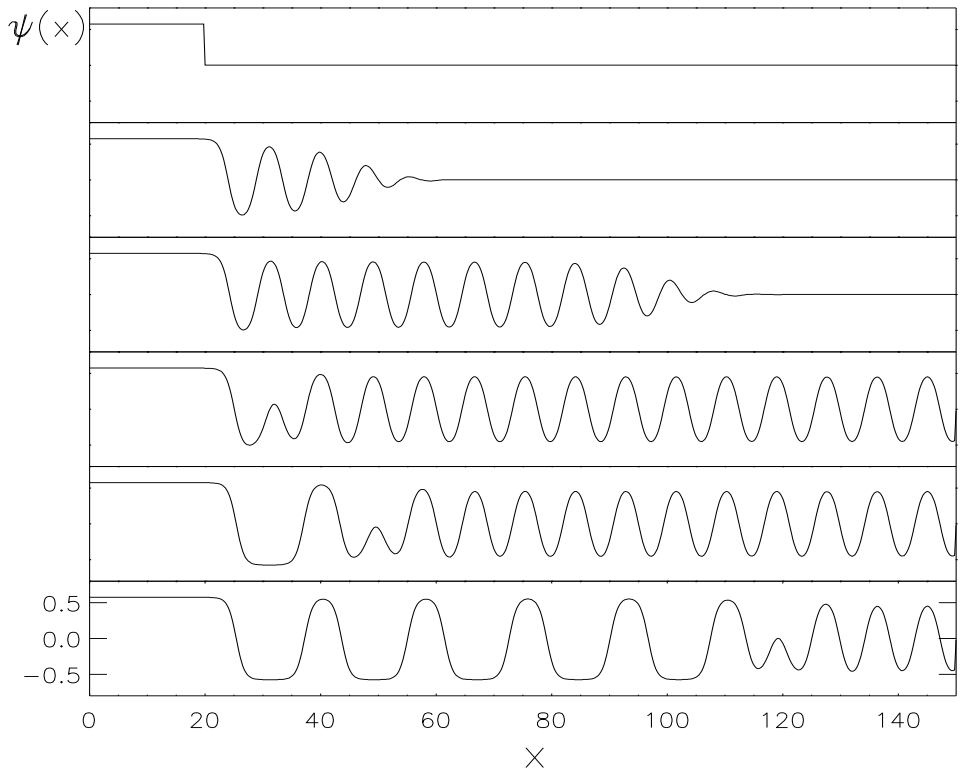

FIG. 2. Graph of $\psi(x, t)$ for $t = 0, 20, 50, 125, 195$, and $455$ for $a = 0.2$.

From the numerical solution we extract the following observations: First, the propagation of the front does not significantly affect the region $x < l$, so that this region remains in the $\psi_+$ stable state. Second, a front advances into the initial homogenous state. Its velocity and the periodicity left behind are well determined by marginal stability theory (Fig. 1). Third and more important, a second front appears and moves into the periodic state left by the first front leaving behind a new periodic state. We interpret this new phenomenon in the following way: The advance of the first front creates in the system a periodic pattern that is very close to one of the periodic solutions of type (c) above. Since such solutions are unstable, they should decay. From the numerical results, the mechanism of decay is exactly the same as the decay of the initial $\psi = 0$ unstable state, that is, a front replaces the unstable state by some 'more stable' state. This interpretation leads naturally to the study of stability properties and of marginally stable fronts propagating into periodic unstable states as relevant for the understanding of the *second-front* phenomenon.

We have been unable to observe a second front in numerical simulations of the EFK model. We think however that this is a consequence of computational limitations and that the second front would appear at sufficiently long times if the numerical scheme is accurate enough. This belief is supported by numerical simulations of EFK in which the initial condition is a periodic state of small wavelength (smaller than the one produced by propagation of the first front) connected to the homogeneous stable state. In this case a front was observed to propagate into the periodic state in very much the same way as in the MCH model. Thus we expect the general reasoning developed here to be of applicability in many transient-pattern forming systems.

## III. FRONTS PROPAGATING INTO PERIODIC UNSTABLE STATES

There are several ways of formulating the marginal stability hypothesis. In all of them the dynamics of the front is analyzed in the leading edge, where the field is small enough so that the equation describing its evolution can be linearized. In the method of van Saarloos, the marginal speed theory is formulated in terms of the evolution equation of a profile with the form of a moving front [10–12]. A second approach is the pinch-point method developed in the context of plasma physics [26,27]. We will use here the method of the steepest descent (or saddle point) [8,13], since it is suitable to be generalized to front propagation into a periodic state, the aim of this section.

We assume that the periodic pattern left by the first front is close to one of the stationary solutions with dominant wavenumber $q = q_i$, that will be denoted by $\psi_{q_i}$. We introduce $\psi(x,t) \equiv \psi_{q_i}(x) + \delta\psi(x,t)$ in the equation defining the model and linearize in $\delta\psi$ to obtain an expression of the form

$$\delta\dot\psi = \left[\mathcal{L} + \mathcal{U}_{q_p}\right]\delta\psi \quad . \tag{3.1}$$

$\mathcal{L}$ is the linear operator giving the linear dispersion relationship $\omega(q)$ corresponding to linearization around the uniform solution $\psi = 0$ and $\mathcal{U}_{q_p}$ stands for the remaining part: a periodic operator of periodicity $q_p$ related to $q_i$. Eq. (3.1) is a linear equation with periodic coefficients whose formal solution is given by Bloch (or Floquet) theory [28,29] in terms of the eigenfunctions and eigenvalues of the linear operator $\mathcal{L} + \mathcal{U}_{q_p}$. According to Bloch theorem, the eigenfunctions $f_k(x)$ of a periodic operator of period $2\pi/q_p$ are products of periodic functions of wavenumber $q_p$ and its harmonics, times plane waves of the wavenumber $k$ which is used to label the eigenfunctions:

$$f_k(x) = e^{ikx} \sum_{r=-\infty}^{\infty} b_r(k) e^{irq_p x} \quad . \tag{3.2}$$

Here $b_r$ are complex amplitudes and $k$ is a continuous index in the infinite size limit. We assume that we are close to this limit, so that sums over $k$ will be approximated by integrals. In addition to $k$, an additional discrete index, labeling eigenvalue branches, is required, but it will be omitted from our formulae to simplify the notation. The integral over $k$ will always denote the sum over the values of $k$ and over the additional discrete index. The Fourier transform of $f_k(x)$, denoted by $\hat{f}_k(q)$ is given by:

$$\hat{f}_k(q) \equiv \int dx\, e^{-iqx} f_k(x) = 2\pi \sum_{r=-\infty}^{\infty} b_r(k)\delta(q - k - rq_p) \tag{3.3}$$

The integration is over the whole system. Given the initial perturbation $\delta\psi_0(x) = \delta\psi(x, t=0)$, the solution of (3.1) reads in terms of these eigenfunctions and the corresponding eigenvalues $\epsilon(k)$:

$$\delta\psi(x,t) = \int_{-\infty}^{+\infty} e^{\epsilon(k)t} f_k(x) C(k) dk \quad , \tag{3.4}$$

where we write the projection of the initial condition onto the Bloch basis as

$$C(k) \equiv \int f_k^\dagger(x)\delta\psi_0(x) dx \tag{3.5}$$

$f_k^\dagger$ is the complex conjugate of $f_k$. Using Parseval's theorem and (3.3), expression (3.5) can be simplified to:

$$C(k) = (2\pi)^{-1} \int \hat{f}_k^\dagger(q)\delta\hat\psi_0(q) dq = \sum_{r=-\infty}^{\infty} b_r^\dagger(k)\delta\hat\psi_0(k + rq_p) \quad . \tag{3.6}$$

Using this result in Eq.(3.4) with Eq. (3.2) we obtain

$$\delta\psi(x,t) = \sum_{r,r'=-\infty}^{\infty} \int_{-\infty}^{+\infty} dk \; b_r(k) b_{r'}^{\dagger}(k) \delta\hat{\psi}_0(k+r'q_p) e^{i(k+rq_p)x+\epsilon(k)t} \quad . \tag{3.7}$$

Let us assume for a moment that only one of the terms in the sum (3.2) is relevant. This will be in fact the case for our particular model, but for completeness we will discuss the general case later. If only one term in the sums over $r$ and $r'$ is relevant, expression (3.7) is formally equivalent to the one used as the starting point in the steepest-descent presentation of marginal stability theory [8,13] for a front moving into an homogeneous state. The criteria for determining the marginal front velocity could then be generalized to our case with the front moving into the periodic state as follows.

The interest is focused in the evolution of the perturbation $\delta\psi(x,t)$, which is supposed to evolve into a front, near its leading edge at large $t$. Restriction to the edge of the front validates the small $\delta\psi$ approximation performed in the linearization leading to (3.1). As the front moves in time, the study of $\delta\psi$ in the leading edge requires using a moving referential frame, that is, taking $z = x - vt$, where $v$ is the velocity of that referential frame. The perturbation is then written as $\delta\psi(z,t)$ and this function has to be calculated at $z \sim 0$ (leading edge) and $t \to \infty$. Precisely for large $t$, the integral can be evaluated by the steepest descent method.

With a single term $r, r' \equiv m$ taken into account and defining $q' = k + mq_p$, Eq. (3.7) in the moving reference frame reduces to

$$\delta\psi(z,t) = \int_{-\infty}^{+\infty} dq' |b_m(q'-mq_p)|^2 \delta\hat{\psi}_0(q') e^{h(q')t + iq'z} \quad , \tag{3.8}$$

we have also introduced $h(q') \equiv iq'v + \epsilon(k = q' - mq_p)$.

According to the steepest descent method one has

$$\delta\psi(z, t \to \infty) \sim e^{h(q'^*)t + iq'^* z} \tag{3.9}$$

Subdominant time-dependencies and unimportant numerical constants have been omitted. $q'^*$ is a saddle point of the function $h(q')$ analytically continued to the complex plane, that is $dh(q')/dq'$ is zero at that point. Using the expression for $h(q')$ one finds that this happens when

$$\left.\frac{d\epsilon(k)}{dk}\right|_{k=q'^*-mq_p} = -iv \tag{3.10}$$

This expression gives the position of the saddle point $q'^*$ in terms of the still undetermined velocity $v$ of the reference frame. As it is assumed a priori that there exists a front moving at a constant speed, we state that when the reference frame moves at the same velocity of the front, $v = v^*$, the perturbation $\delta\psi$ should remain finite and bounded in time in the vicinity of the leading edge ($z \sim 0$). From (3.9) this gives the additional requirement

$$\text{Re}[h(q'^*)] = 0 \tag{3.11}$$

From this equation and (3.10) with $v = v^*$ one can determine the velocity of the front $v^*$ and the complex number $q'^*$ locating the saddle point. These equations are formally equivalent to (1.1), obtained for a front moving into an homogeneous state. The interpretation of $q'^*$ is the same as in such case: Eq.(3.9) shows that the real part of $q'^*$ gives the periodicity of $\delta\psi$ at the edge of the front, and its imaginary part characterize its steepness. The difference with the homogeneous case lies in the different eigenvalue spectrum $\epsilon(k)$.

Generalization of the above results to the case in which all the terms in the sum (3.2) are retained is now immediate. It is enough to estimate with the saddle point method each term in the sum (3.7). An equation identical to (3.10) is obtained for each value of $r$, which replaces $m$. There is now one value of $q'^*$ for each $r$, but (3.10) shows that all of them lead to the same $k^*$. After some algebra we obtain

$$\delta\psi(x,t) \sim e^{\epsilon(k^*)t} f_{k^*}(x) \quad , \tag{3.12}$$

where subdominant time dependencies and factors independent of $x$ and $t$ have been omitted. Using (3.2), and expressing (3.12) in the moving reference frame, $z = x - vt$, the condition of steady envelope of the front near the leading edge leads to the following set of equations (one for every $r$):

$$\text{Re}\left[\epsilon(k^*) + i(k^* + rq_p)v\right] = 0 \tag{3.13}$$

Since $q_p$ is a real number all these equations are in fact the same Eq.(3.11). Then (3.10) and (3.11) are valid also in this situation in which several modes in (3.2) are relevant. The only difference is that the state growing out of the unstable initial condition (see (3.12)) is not a single Fourier mode but a Bloch eigenfunction.

The wavelength $\lambda$ of the periodic pattern left behind by the moving front can be calculated following the prescription of Dee and Langer [8,10–12]. The main idea is to assume that the oscillations created at the leading edge by the linear instability will become quenched by the nonlinearities but their periodicity not modified. It is also necessary that one of the terms in Eq. (3.2) (say $r = m$) is the most important, so that Eq. (3.9) applies. In the moving frame of speed $v^*$, linear theory predicts that the leading edge (3.9) oscillates at a frequency such that a number of nodes $\Phi$ is created in the unit of time, with $\Phi = \pi^{-1} \left( \text{Im}\left[\epsilon(k^*)\right] + v^* \text{Re}\left[q'^*\right] \right)$. Behind the front, where the pattern has a wavelength $\lambda$, the flux of nodes passing in the unit of time through a point fixed in the moving frame is $2v^*/\lambda$. From this we get

$$\lambda^{-1} = \frac{1}{2\pi} \left( \frac{\text{Im}\left[\epsilon(k^*)\right]}{v^*} + \text{Re}\left[q'^*\right] \right) , \qquad (3.14)$$

which is a generalization of (1.2) but now $\lambda$ is determined by the two different wavenumbers $k^*$ and $q'^*$. The hypothesis behind this formula are that no nodes are created nor destroyed far from the leading edge, and that every node that linear analysis predicts to be created has to be really created. These assumptions work very well in the cases studied by van Saarloos and also in our case, but it should be remarked that they are not general and in some cases they seem to be violated, as for example in the case studied by Dee [21].

## IV. SECONDARY FRONTS IN THE MCH MODEL

The periodic pattern generated by the first front contains a dominant wavenumber $q_i$. Our basic hypothesis to explain the second front phenomenon is that the pattern can be well represented by one of the stationary solutions of type (c) with such fundamental wavenumber. We study its stability with respect to new propagating fronts. To apply the formalism of the previous section, we first need the expression for $\psi_q(x)$, the solution associated with a given wavenumber $q$. Then the spectrum of the operator $\mathcal{L} + \mathcal{U}_{q_p}$ should be calculated and analytically prolongated to the complex plane. The solution of (3.10) and (3.11) gives then the velocity of the propagating front. None of these steps can be done exactly in an analytical way for the MCH model. They can in principle be fully done numerically, but since our aim is to explain some phenomena that have been already observed numerically, it would be preferable obtaining some approximate analytical expressions. We can find such approximation by working in the situation in which the wavenumber of the initial unstable configuration is close to $\sqrt{c}$. In this case the amplitude of $\psi_q$ is small, so that the associated $\mathcal{U}_{q_p}$ can be considered a small perturbation to $\mathcal{L}$. In addition, only the fundamental harmonic of $q$ is important in this limit, a fact that simplifies calculations. By looking at Fig. 1, we see that the pattern created by the first front never has a periodicity $q_i = \sqrt{c}$ (remember $c \equiv 1$), but for sufficiently small $a$ the approximation makes sense, since $q_i$ comes close to 1.

To begin with, an approximate stationary solution of Eq.(2.2) (Duffing equation) is obtained by standard methods [30]:

$$\psi_{q_i}(x) \approx \left(4(c - q_i^2)/3b\right)^{1/2} \sin(q_i x) \qquad (4.1)$$

This solution is valid for $q_i \lesssim \sqrt{c}$, so that $\psi_{q_i}$ is small. Substitution of $\psi = \psi_{q_i} + \delta\psi$ into equation (2.2) gives

$$\delta\dot{\psi} = (a - \partial_x^2)(c + \partial_x^2 - 3b\,\psi_{q_i}^2)\delta\psi . \qquad (4.2)$$

Comparing with (3.1), we identify $\mathcal{U}_{q_p} = -3b(a - \partial_x^2)\psi_{q_i}^2$, which is small for $q_i \lesssim \sqrt{c}$ and has a periodicity $q_p = 2q_i$. Writing

$$\psi_{q_i}^2(x) = \sum_{l=-1,0,1} A_l e^{2il q_i x} \qquad (4.3)$$

where $A_0 = 2(c - q_i^2)/(6b)$, and $A_1 = A_{-1} = -A_0/2$, the eigenvalue problem for the equation (3.1) can be written as

$$\epsilon(k)b_l(k) = \omega(q_l)b_l(k) - \eta(q_l) \sum_{m=-1}^{1} A_m b_{l-m}(k) \quad ; \quad l = -\infty, \cdots, +\infty \qquad (4.4)$$

with $\omega(q_l) = (a + q_l^2)(1 - q_l^2)$ and $\eta(q_l) = 3b(a + q_l^2)$, where $q_l = k + 2l q_i$. This expression represents, for every $k$, an infinite number of coupled equations for $\{b_l\}$.

Several methods are known for calculating the spectrum of periodic operators approximately. Here we use the simplest weak coupling approximation in which $\mathcal{U}_{q_p}$ is treated as a small perturbation of $\mathcal{L}$ [28]. The largest effect of the perturbation occurs at the boundary of the Brillouin zone $k = \pm q_i$ where the degeneracy of the free spectrum $\omega(k) = \omega(k - q_p/2)$ is removed creating a gap. However for $k$ close to $q_i$ the two dominant free eigenfunctions are still $e^{\pm i q_i x}$ which correspond to $r = 0, -1$ in (3.2). Near the Brillouin boundary (4.4) can then be reduced to a $2 \times 2$ matrix equation setting $b_r = 0, r \neq 0, -1$. The resulting equation have nontrivial solution for

$$\epsilon^{\pm}(k) = \left(\frac{1}{2}\right)\left(-A_0(u + u_1) + (w + w_1)\right.$$
$$\left. \pm \sqrt{(w - w_1)^2 + A_0^2(u^2 - u u_1 + u_1^2) + 2A_0(u_1 w - u w + u w_1 - u_1 w_1)}\right) , \qquad (4.5)$$

where $w = \omega(k)$, $w_1 = \omega(k - 2q_i)$, $u = \eta(k)$ and $u_1 = \eta(k - 2q_i)$. It can be shown that $\epsilon^+(k) \geq 0$ for all $k$ while $\epsilon^-(k)$ is negative. So $\epsilon^+(k)$ gives the energy of the upper band an it is the required amplifying factor to be used in the integral (3.4). Expression (4.5) is only valid for $k \approx q_i$. In fact, if $k = q_i$ gives $\epsilon^+ = 0$, which is the exact eigenvalue associated to a uniform translation of the pattern. However far from $k \approx q_i$ it lacks some properties of the exact spectrum, as for example the symmetry $k \to -k$.

The marginal speed of the front has to be obtained now by substitution of $\epsilon^+(k)$, with $k$ taken to be complex, into Eq.(3.10) and (3.11). The resulting set of three real algebraic equations cannot be solved analytically. The method used to find the solution is as follows: the fourth order Taylor expansion of $\epsilon^+(k)$ around a point $k = k_t$ is determined. Then, the derivatives involved in Eq.(3.10) are analytically calculated and the resulting set of equations numerically solved for $v^*$ and $k^*$. As the Taylor expansion only fits the original function near a small neighborhood of $k_t$, an iterative selfconsistent method is applied: $k_t$ is initially taken as the point where $\epsilon^+(k)$ for $k$ real takes its maximum value. A first approximation for $v^*$ and $k^*$ is numerically obtained and a new Taylor expansion of $\epsilon^+(k)$ is calculated around a new $k_t$ equal to the real part of the just obtained $k^*$. The process is repeated until the solutions converge. The final solution turns out to be independent of the initial point $k_t$ provided that it is not too far from the maximum of $\epsilon(k)$ with $k$ real. Once the solutions $v^*$ and $k^*$ were obtained by this method, they were proved to verify the relevant set of equations.

Within the approximation used to solve (4.4) it is found that as $k$ moves away from the Brillouin boundary, $b_{-1}$ is the largest amplitude in (3.2) for the upper band of the spectrum ($\epsilon^+(k)$). Then, as $q'^* = k^* - 2q_i$ is the most important wavenumber of the function $f_{k^*}(x)$ the use of Eq. (3.14) gives the periodicity of the pattern left behind.

In the next section, the velocity of the second front calculated from this theory as a function of $q_i$ and also the wavenumber of the pattern left behind will be presented and compared with numerical results from the direct integration of (2.2).

## V. NUMERICAL RESULTS AND COMPARISON WITH THEORY

Figure 3 shows space-time plots of $\psi(x, t)$ obtained from a numerical solution of (2.2) in gray levels and for several values of $a$. The horizontal axis represents space, and time runs along the vertical axis. White corresponds to regions with $\psi$ near the value of the stationary stable solution $1/\sqrt{b}$ and black to those with the value near $-1/\sqrt{b}$. The location of the fronts is clear and the measured slopes determine their velocities.

For the first front moving into the homogeneous phase, the velocity obtained fits the result obtained with the marginal stability theory (see Fig. 1). Each value of the parameter $a$ determines through (3.14) a periodicity which is in agreement with the numerical solution. Such periodicity identifies a $q_i$ which is used as the initial wavenumber to apply the general results of IV for the second front. The values obtained from the theory and from the numerical solution for the second front are shown in Fig. 4a for several values of $a$. The mean periodicity left behind the second front is shown in Fig. 4b together with those obtained from Eq. (3.14). The error bars on the theoretical values are obtained when the value of $q_i$ assumed for the first front is modified by $\pm 0.01$ from the value predicted from marginal stability theory. This intends to take into account the fact that due to system size the wavenumbers in the numerical simulations are discrete and can then differ in this amount from the theoretical predictions. It is seen that this small indeterminacy on the value of $q_i$ can have quite a large effect in the final states. In addition this also suggest that small deviations from perfect periodicity in the state left by the first front can induce relevant changes in the state left by the second front.

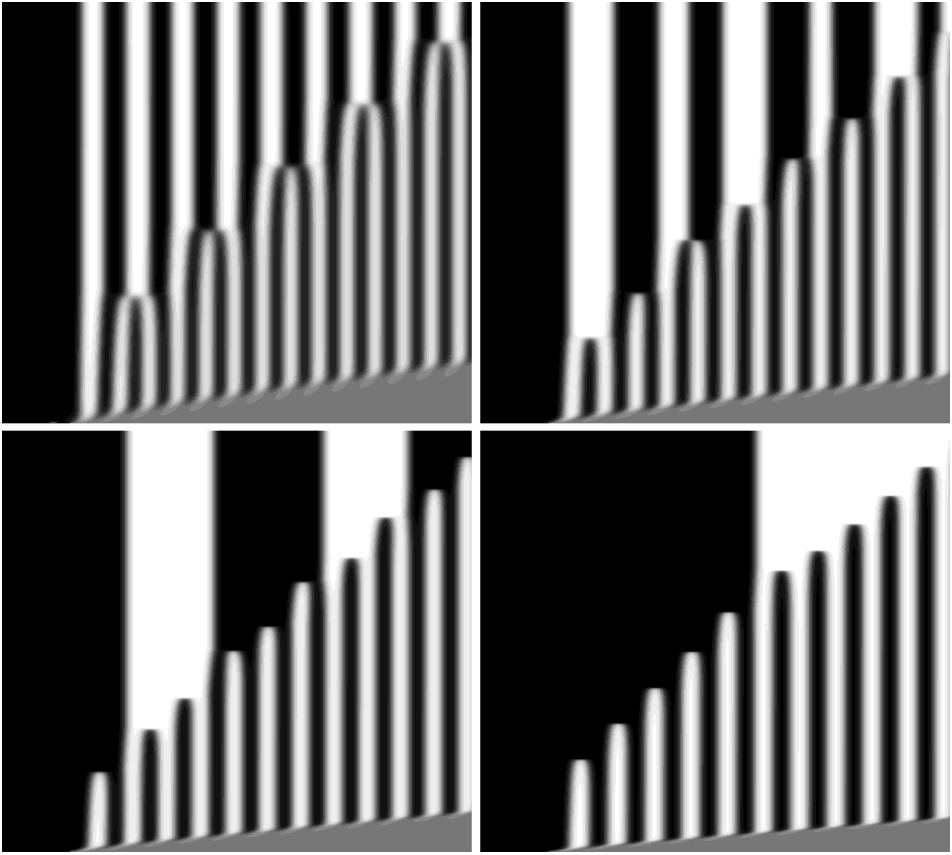

FIG. 3. Space-time plots of $\psi(x,t)$ in gray levels, for $a = 0.0, 0.3, 0.6$ and $0.9$. Space is represented along the horizontal axis (system size: 137.5) and time runs along the vertical one from 0 to 575. White and black correspond to values of $\psi(x,t)$ close to plus and minus $1/\sqrt{b}$, the value of the homogeneous stationary solutions.

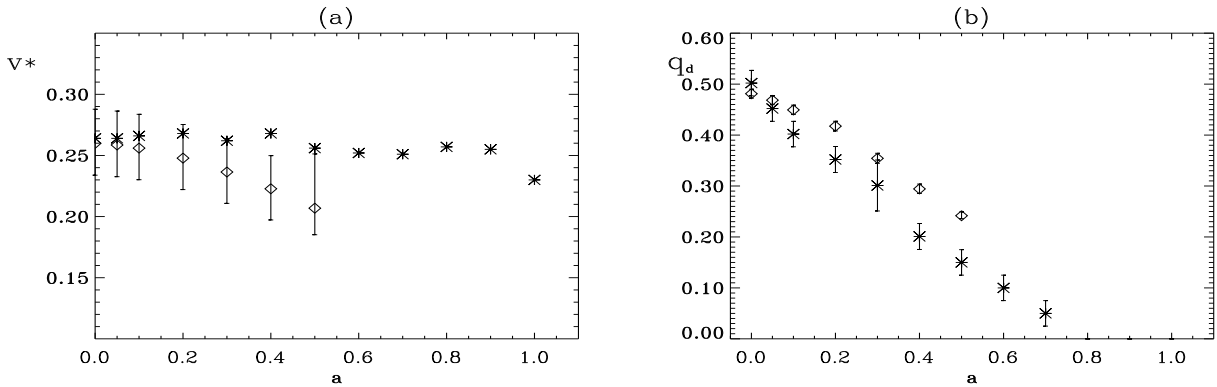

FIG. 4. (a) Speed of the second front for different values of the parameter $a$, obtained by using the marginal stability theory ($\diamond$) and obtained from the numerical simulations ($*$). The vertical bar is a tolerance band (see text). (b) Comparison of the dominant wavenumber left behind the second front.

Fig. 4a and Fig. 4b indicate that, as expected, our approximate analytical calculation is accurate for small values of $a$. In fact, the periodicity $q_i$ is closest to $\sqrt{c}$ for $a = 0$ and the weak coupling calculation of the spectrum $\epsilon(k)$ is only meaningful for $q_i \lesssim \sqrt{c}$. The validity of the marginal stability hypothesis for the general case studied in section III of a front propagating into a periodic unstable state, can be checked applying the result of section IV for values of $q_i$ for which (4.5) is justified. These values do not generally correspond to a pattern created by a first front for a particular value of $a$. To this end, we have studied the appearance of a front for $a = 0$, and starting from the exact stationary periodic solution, instead of letting it be created by the first front. The stationary state was generated numerically taking $\psi(x,0)$ from (4.1), and then use (2.2) until $\psi$ becomes steady with the growth of the harmonics modes. Then, a small part of this solution is replaced by $\psi_+$ and we observe the evolution of the resulting configuration. For $q_i > 0.9$ the pattern is already very unstable and numerically it decays by roll annihilation before the appearance of the second front. On the other hand, for $q_i < 0.6$ the velocity of the second front tends to be so small that the computer time needed to observe it becomes prohibitive. For intermediate $q_i$ we show in Fig. 5, the velocity of the front and the wavenumber of the periodic pattern left behind the front (determined from the average wavelength of the pattern). The agreement between the theoretical values and the numerical ones is quite good, taking into account that the calculations of $\epsilon(k)$ is accurate for $q_i \to \sqrt{c} \equiv 1$.

The wavelength of the pattern left by the second front becomes larger as the value of $a$ is increased. Our numerical results indicate the possibility of the divergence of the wavelength for a finite value of $a$, in the same way than the wavelength of the first front diverges for $a \leq c(7 + 4\sqrt{3})$. However the finite size of our simulations makes it difficult to study further such a possible divergence.

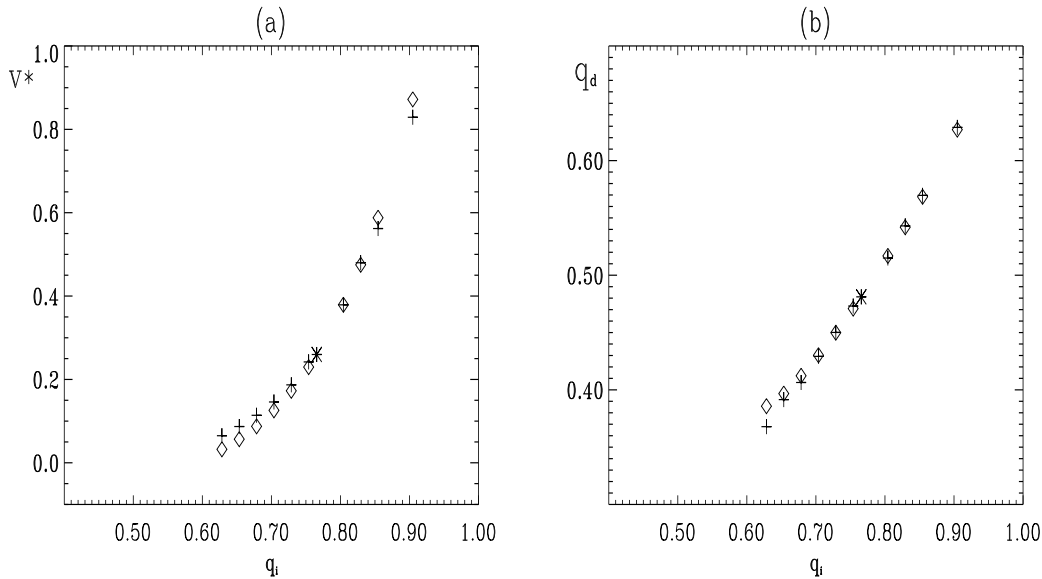

FIG. 5. Velocity of the second front and wavenumber of the structure created when the front propagates into stationary states $\psi_{q_i}(x)$ with different dominant wavenumber ($a = 0$). The point represented by $*$ corresponds to the value of $q_i$ obtained in the wake of the first front, for $a = 0$.

The velocity of the second front is near one order of magnitude smaller than that of the first front. This means that the region occupied by the periodic pattern left by the first front will become larger as time goes on. Hence, it is expected that the decay in this region, far enough from the boundary condition, will occur finally through the local mechanism of roll destruction analyzed in [23], instead than through the second front mechanism. This is enhanced by the fact that the pattern left by the first front is not exactly periodic. In fact marginal stability theory is well founded only when the perturbation around the initial unstable state is localized in space. This assures that the dependence on the initial condition present but not explicitly written in (3.9) and (3.12) plays no role in the selection process. The difference between the pattern created by the first front and the closest periodic stationary solution is not localized and the mode of decay at very long times can be different from the front mechanism. Hence, as observed, the second front can only advance a finite (yet undetermined) distance before the pattern left by the first front decays by a bulk mechanism.

This mode of spontaneous decay of the pattern, that is, unaffected by the boundary, seems to be also the general way of evolution of the structure left by the second front. In fact this pattern has a Fourier spectrum broader than the produced by the first front, and even the local wavenumber is slightly different from place to place. However, a very interesting case has been observed for $a = 0.2$. The pattern left by the second front decays towards the homogeneous state through the appearance of a third front. This is shown in the space-time plot of Fig. 6. The root for this behavior seems to be the fact that the wavenumber of the second structure is fairly twice that of the first and this leads to a second structure with a very well-defined periodicity. For other values of $a$ where there was an integer ratio between both wavenumbers, the second pattern has a wavenumber so small that we have not been able to observe numerically the decay of the structure in a reasonable computer time. In fact, when the wavenumber of the pattern tends to zero the main driving force for decay is wall interaction which leads to a very slow logarithmic evolution.

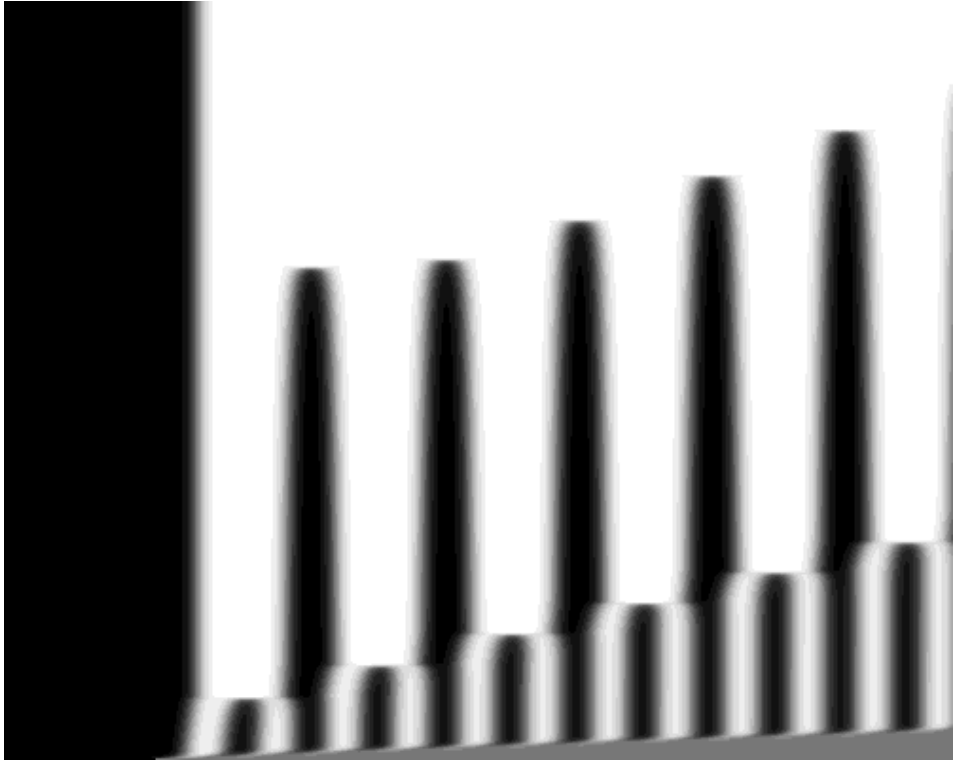

FIG. 6. Space-time plot of $\psi(x,t)$ as in Fig. (3) for $a$ is 0.2 (system size: 125.7 ; time from 0 to 1600. The decay of the unstable state $\psi = 0$ to $\psi = 1/\sqrt{b}$ through a sequence of three fronts is clearly observed.

## VI. REMARKS AND OPEN QUESTIONS

The main conclusion of our work is the existence of a second front mechanism for the decay of unstable periodic pattern created by a first front in the MCH and CH models. This second front is well explained by marginal stability theory around a periodic state. As commented at the end of section II, we expect the second front phenomena to be a general feature of transient pattern forming models. The question of the observability of this phenomena in real experiments is related to the same question for the first front: due to thermal and environmental noise it is experimentally difficult to maintain a sample in a unstable state long enough to observe the advance of propagating fronts. As a consequence there are many experiments on front propagation into metastable states, but examples of front propagation into unstable states are scarce [15,17]. Nevertheless, the transient patterns observed during the Freédericsz transition in nematics [31] appear to be quite long lived, so that this system which is generally described by (2.2) [23,20] is an obvious candidate to look for the second front.

Our theoretical approach has been intentionally restricted here to the situation in which some analytical or semi-analytical results could be obtained. A number of open questions which could be addressed by a more detailed and extensive numerical study include the possible divergence of the wavelength of the pattern behind the second front at a critical value of $a$, the interplay between front propagation and bulk decay and the conditions for observability of third and subsequent fronts.

Finally it should be stressed that marginal stability theory in its present state is just an heuristic prescription. It would be desirable to establish rigorously its range of applicability. Such rigorous analysis has been only performed so far for parabolic equations [7].

## ACKNOWLEDGMENTS

This work has been partially supported by Dirección General de Investigación Científica y Técnica (DGICyT, Spain), under contract PB92-0046-C02-02. R.M. also acknowledges partial support from the Programa de Desarrollo de las Ciencias Básicas (PEDECIBA, Uruguay), the Consejo Nacional de Investigaciones Científicas Y Técnicas (CONICYT, Uruguay) and the Programa de Cooperación con Iberoamérica (ICI, Spain).